\documentclass[prl,twocolumn,superscriptaddress,showpacs,floats,floatfix]{revtex4}
\usepackage{epsfig,dcolumn,amsmath,latexsym}
\begin{document}

\title{Performance limits of graphene-ribbon-based field effect transistors}

\preprint{1}

\author{F. Mu\~noz-Rojas}
\author{J. Fern\'andez-Rossier}
\affiliation{Departamento de F\'\i sica Aplicada, Universidad de Alicante, 
San Vicente del Raspeig, E-03690 Alicante, Spain.}
\author{L. Brey}
\affiliation{Instituto de Ciencia de Materiales de Madrid, Consejo Superior de Investigaciones Cient\'\i ficas, 
E-28049 Cantoblanco, Spain}
\author{J.\ J.\ Palacios}
\affiliation{Departamento de F\'\i sica Aplicada, Universidad
de Alicante, San Vicente del Raspeig, E-03690 Alicante, Spain.}
\affiliation{Instituto de Ciencia de Materiales de Madrid, Consejo Superior de Investigaciones Cient\'\i ficas, 
E-28049 Cantoblanco, Spain}

\date{\today}

\begin{abstract}
The performance of field effect transistors based on an single graphene ribbon with a constriction 
and a single back gate are studied with the help of atomistic models. It is shown how
this scheme, unlike that of traditional
carbon-nanotube-based transistors, reduces the importance of the specifics of the chemical 
bonding to the metallic electrodes in favor of the carbon-based part of device. 
The ultimate performance limits are here studied for various constriction and metal-ribbon
contact models. In particular we show that, even for poorly contacting metals, properly taylored constrictions can 
give promising values for both the on-conductance and the subthreshold swing.
\end{abstract}


\maketitle

A number of factors determining the performance of carbon nanotube field effect transistors (CNTFET's) still
remain to be mastered before these promising devices can compete with current Si-based transistors.
One of these factors, probably the most important one, is the Schottky barrier formed
at the interface between the source or drain metallic electrodes and the carbon 
nanotube\cite{Heinze:prl:02,Javey:nature:03,Yang:apl:05,Chen:nl:05}.
Current understanding attributes the different observed performances of CNTFET's to
the different work function values of the metals used as electrodes. Furthermore, even for the same metal, a 
disparity of behaviors have been observed due to the difficulty in controlling the Schottky barrier, i.e., the
resulting position of the Fermi energy with respect to the  valence and conduction bands of 
the nanotube as a consequence of different fabrication schemes. These translate 
into a variety of barrier heights for holes or electrons and a variety of on-currents and threshold
voltages for either voltage gate polarity\cite{Heinze:prl:02,Javey:nature:03,Yang:apl:05,Chen:nl:05}.  

\begin{figure}
\includegraphics[width=\linewidth]{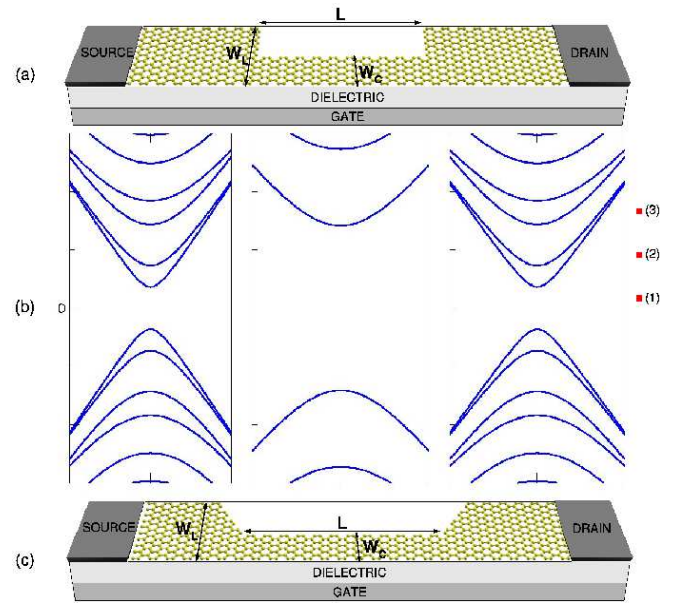}
\caption{ (Color online). Schematic view of a graphene ribbon field effect transistor with a square (a) and tapered (c)
constriction placed on one side.
(b) Example of band structures corresponding to the three regions: Left lead, right lead, and 
constriction. The horizontal lines denote different Fermi levels corresponding to different conduction situations: 
(1) at the charge neutrality point, (2) off state, and (3) on state.} 
\label{GRFET}
\end{figure}

A way to get around the lack of reproducibility of current-voltage and transfer characteristics of
CNTFET's is to avoid the use of metallic electrodes
close to the semiconducting channel. This has been attempted by integrating ohmic contacts in the 
carbon nanotube through a combination of back and top gates\cite{Javey:nl:04} which allow for an independent control 
of the density on different regions of the nanotube. This scheme allows to create metallic and semiconducting regions
on a single nanotube. Band-gap
engineering of graphene ribbons\cite{Nakada:prb:96,Brey:prb:06}, on the other hand, offer an alternative strategy to the 
multiple-gate scheme\cite{Han:prl:07,Pablo}.  A single graphene ribbon with armchair edges and 
a narrow constriction in the middle [see Figs. \ref{GRFET}(a) and (c)]
can present three regions with different band gaps which can be doped differently by a single back gate 
[see Fig. \ref{GRFET}(b)]. For constriction widths with a number a dimer lines $N \neq 3M-1$,  with $M$ an integer
smaller than $\approx 500$ the constriction presents a transport gap at room temperature\cite{Son:prl:06}
and acts as the semiconductor active channel in standard transistors. 
Since the gap scales inversely with the width of the ribbon, the wide sections of the ribbon
will always present a  zero or vanishingly small gap. These can easily behave as metallic eletrodes or leads because
of finite temperature or
as they are driven out of the charge neutrality point by the action of the back gate voltage, $V_{\rm G}$.
The same gate voltage controls the on-off state by
bringing the Fermi level into the gap of the constriction or out of it [see Fig. \ref{GRFET}(b)] 

Theoretically, this system should behave as an ambipolar transistor where the threshold voltage, $V_{\rm th}$, is
determined solely by the width of the constriction\cite{Fernandez-Rossier:prb:07}.
The on-conductance, $G_{\rm on} = dI/dV|_{V_{\rm G} > V_{\rm th}}$,
and the subthreshold swing, $S=(d\log{I}/dV_{\rm G})^{-1}$, magnitudes that determine the transistor performace, 
are also controled by the capacitive couplings and by the {\it atomic structure of the lead-constriction contact}.
The ultimate performace limits of graphene ribbon field effect transistors 
(GRFET's) are here established for two contact models. The first,
{\it square},  where the crystallographic orientation of one edge changes by 90$^{\circ}$ at the contact 
[see Fig. \ref{GRFET}(a)]. The second, {\it tapered},
where the lead narrows down progressively until reaching the constriction width. This is achieved in practice by a single
change in the crystallographic orientation of one of the  edges of 60$^{\circ}$ [see Fig. \ref{GRFET}(c)]. 
In this case the armchair edge is never interrupted, except for the change in the crystallographic orientation. 
The constriction is always placed laterally, i.e., on one side of the ribbon.
This choice can be justified on the basis that, in order to fabricate a constriction with reduced disorder, it is 
desirable to leave untouched one edge of the initial graphene layer or 
wide ribbon and, e.g., etch away only the other edge 
(constrictions created by etching both edges have already been fabricated\cite{Han:prl:07}). Our proposed
GRFET could simplify the fabrication process and might improve the final performance.
A second justification will be mentioned when discussing the results.

{\it Methodology.--} We compute  the conductance of the constriction, $G_{\rm C}$, using the Landauer formalism 
which assumes coherent transport across the constriction.   The electronic structure is
calculated in the standard one-orbital tight-binding 
model\cite{Nakada:prb:96,Fujita:jpsj:96,Brey:prb:06,Munoz-Rojas:prb:06,Fernandez-Rossier:prb:07} 
which has been shown to reproduce the low-energy physics of ribbons whose edges are 
saturated with hydrogen.  The scattering problem is solved
using the Green's function approach. This involves (i) the calculation of the Green's function 
projected on the constriction and on part of the left and right ribbon, (ii) the calculation of the 
self-energies for the semiinfinite ribbon leads by iterative solution of the Dyson equation\cite{Munoz-Rojas:prb:06}, 
and (iii) the subsequent evaluation of the transmission probability,
$T_{\rm C}$, using the Caroli expression\cite{Caroli:jphysc:71}. The details can be found in 
Ref. \onlinecite{Munoz-Rojas:prb:06}. In order to simplify the discussion, we assume first
semiinfinite graphene ribbons on both sides. The role played by the metallic electrodes is considered
at the end, although one can anticipate that 
they should not affect the results as long as $W_{\rm L}\gg W_{\rm C}$, where $W_{\rm L}$ is the width of the bulk ribbon
and $W_{\rm C}$ is the width of the constriction at its narrowest point.

\begin{figure}
\includegraphics[width=0.9\linewidth]{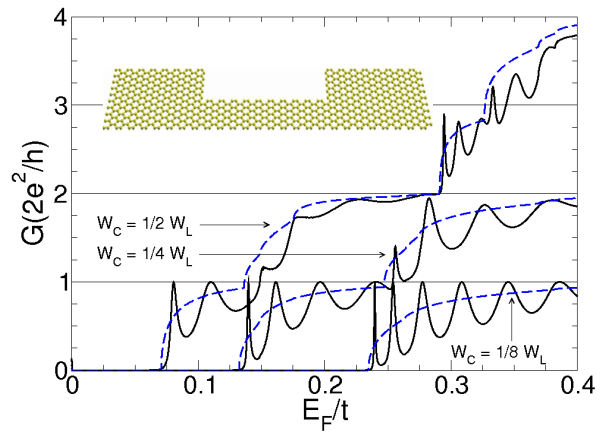}
\caption{(Color online). Solid lines correspond to the conductance of an $L\approx 10$nm constriction for different widths
in the case of the square contact model shown in the inset. The width of 
the lead is $W_{\rm L}=5.8$ nm. In dashed lines the conductance for a single interface is shown.}
\label{90}
\end{figure}

\begin{figure}
\includegraphics[width=0.9\linewidth]{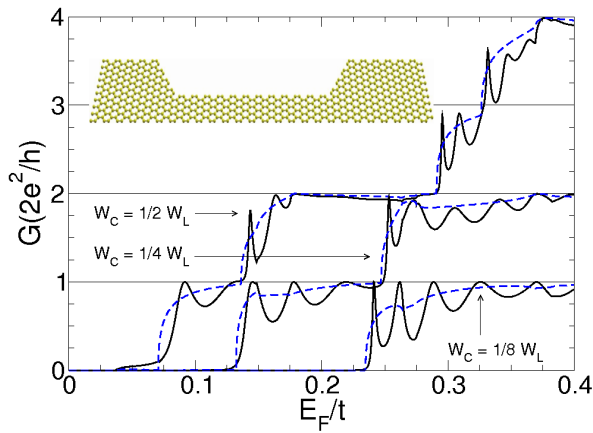}
\caption{(Color online). Same as in Fig.\ref{90}, but for 
the tapered contact model shown in the inset.}
\label{60}
\end{figure}

{\it Zero-temperature results.--}
We begin by showing in Figs. \ref{90} and \ref{60} 
the zero-temperature $G_{\rm C}$ for a constriction with length $L \approx 10$  nm
and $W_{\rm C}=$1/2, 1/4, and 1/8 $W_{\rm L}$, for $W_{\rm L}$=5.8 nm. 
For both types of contacts we obtain the expected step-wise increase of the conductance as a function of
$E_{\rm F}$ (in units of the hopping parameter $t$), 
associated with the increase in the number of bands crossing the Fermi level in
the constriction. The steps come in pairs, reflecting the band structure of semiconducting armchair ribbons [see
Fig. \ref{GRFET}(b)]. On top of the steps we obtain strong Fabry-Perot-like oscillations 
as a consequence of the finite reflection at the constriction-lead contacts. The periodicity of these oscillations 
is consistent with
the length and with the dispersion relation in the constriction.  The amplitude of the oscillations, on the other hand,
depends on how abrupt the contact is. For the square contact 
the amplitude of the oscillations is larger than for the tapered one,  as expected.
In order to separate the contribution of the scattering at the interfaces from
the quantum interference effects, we have also computed the transmission of a single 
interface (dashed lines in Figs. \ref{90} and \ref{60}).
In all cases the transmission steps saturate to their quantum limit very slowly.
For both contact types, as the width of the constriction decreases,
the transmission worsens as a consequence of the increasing mismatch between lead and constriction wave functions.
The increasing reflection on decreasing $W_{\rm C}/W_{\rm L}$ 
is somewhat damped for the tapered contacts, but still clearly visible.
When $W_{\rm C}/W_{\rm L} \rightarrow 0$ quasilocalized states form in the 
constriction as a result of a significant loss in the
transparency of the contacts. This situation resembles the formation of quantum dots in poorly contacted 
carbon nanotubes\cite{Bockrath:nature:99}, but without metallic contacts.
In general one can anticipate that constrictions with tapered contacts to the leads should 
perform better than their square counterparts for both $G_{\rm on}$ and $S$, although the overall performace diminishes as 
the constriction narrows down in both models.

\begin{figure}
\includegraphics[width=0.9\linewidth]{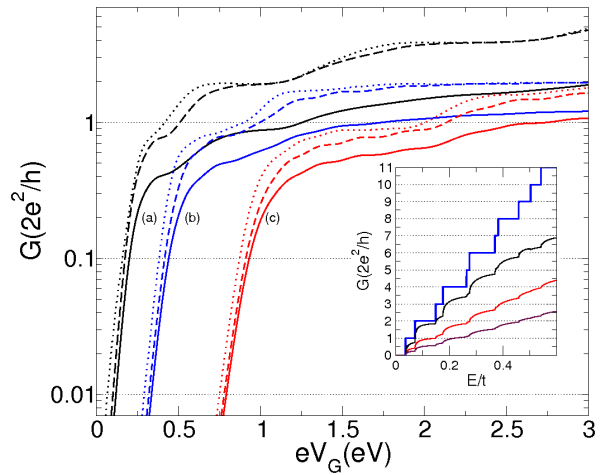}
\caption{(Color online).
Logarithmic plot of the room-temperature conductance for the square (dashed lines) and tapered (dotted lines)
constrictions shown in Figs. 1(a) and 1(c) for different constriction widths $W_{\rm C}/W_{\rm L}$=1/2 (a), 
1/4 (b), and 1/8 (c), with $W_{\rm L}=5.8$ nm. For other parameters see text.  Solid
lines denote the total conductance of the square constriction including metal electrodes with on-site energy
$\epsilon=5t$.  The small panel shows the conductance of the metal-graphene interface
for $\epsilon=0t,\, 5t,$ and $10t$ (top to bottom).}
\label{T300}
\end{figure}

{\it Room-temperature results.--}
We now turn our attention to the actual room-temperature performance of GRFET's.
Although the widths and lengths considered ($\leq 10 $nm) are still difficult to achieve with present
lithographic techniques\cite{Han:prl:07}, the results can be easily extrapolated to more realistic constrictions. 
The ones considered here behave as intrinsic semiconductors at room temperature
since their  gaps are in the range $\approx 0.1-1$ eV for a typical value of $t=2.7$ eV, which 
is much larger than $kT \approx 25 $ meV at room temperature.
Figure \ref{T300} shows a logarithmic plot of $G(V_{\rm G})$ for the all the cases previously studied. 
In all cases the conductance oscillations have disappeared, smeared out by temperature.
The exact relation between $E_{\rm F}$ and $V_{\rm G}$
depends on the actual capacitive couplings between graphene, gate, source, and drain electrodes. 
Modeling these couplings is beyond the scope of this work\cite{Liang:07}. However,
for wide and infinite ribbons where the source and drain 
electrodes are infinitely far away from each other an approximate analytic expression
between $E_{\rm F}$ and $V_{\rm G}$ can be obtained\cite{Fernandez-Rossier:prb:07}: 
\begin{equation}
eV_G =\frac{e^2 d}{\epsilon} \frac{16\beta}{3 t^2a^2} E_{\rm F}^2 + E_{\rm F},
\label{Vgate2D}
\end{equation}
where $a$ is the graphene lattice constant, $\epsilon$ is the dielectric constant, and
$\beta$ is the capacitance ratio between graphene and graphene ribbon\cite{Fernandez-Rossier:prb:07}.
Here we have considered a realistic 
situation that reduces the classical electrostatic capacitance: $d=7$ nm, $\epsilon = 47$.
Based on self-consistent calculations using the same
methodology in Ref. \onlinecite{Fernandez-Rossier:prb:07}, we estimate $\beta$ to be $\approx 0.16$.

We can now extract the respective values of $S$ from the logarithmic plots. (Note that these values can
be additionally taylored through $d$, $\epsilon$, and $\beta$.) 
While, for the wider constrictions, $S$ is larger for the tapered contacts  (90 vs. 74  meV/decade), 
these contacts outperform the square ones for the narrowest constrictions (125 vs. 132 meV/decade). In any case,
$S$ increases as the constriction narrows down.  Also,
the combined action of the finite reflection at the interfaces, finite temperature, and tunneling
prevents $G_{\rm on}$ from reaching the quantum limit $2e^2/h$ for $V>V_{\rm th}$ before the next channel
opens up. 

{\it The role of metal contacts.--}
Finally, to approach a real experimental situation, we consider the role played by the source and drain
metallic electrodes which are needed to connect the ribbon to the external cicuitry. 
At room temperature, when the metallic electrodes are sufficiently far from the constriction,
electrons may loose phase coherence before reaching it.
In this limit we can consider the total conductance to be given by
\begin{equation}
G_{\rm T}= (R_{\rm C} + R_{\rm M}^{\rm S}+ R_{\rm M}^{\rm D}+ R_{\rm Q})^{-1},
\end{equation}
where $R_{\rm C}$ is the so-called four-terminal resistance of the constriction,
and $R_{\rm M}^{\rm S}$ and $R_{\rm M}^{\rm D}$ are the four-terminal resistances (hereon considered equal) 
due to the contact with the source and drain metal electrode, respectively,
\begin{equation}
R_{\rm C(M)}=\frac{h}{2e^2}\frac{N-T_{\rm C(M)}}{NT_{\rm C(M)}},
\end{equation}
A new transmission function, $T_{\rm M}(E_{\rm F})\leq N(E_{\rm F})$, has been introduced to account for 
the scattering at the interface between the metal and ribbon.  Finally,  $R_{\rm Q}=\frac{h}{2e^2}\frac{1}{N}$ is the
intrinsic quantum resistance of the ribbon. In this approximation the total conductance is given by
\begin{equation}
\label{total}
G_{\rm T}=\frac{2e^2}{h}\left(\frac{2}{T_{\rm M}}+\frac{1}{T_{\rm C}}-\frac{2}{N} \right)^{-1}.
\end{equation}
From Eq. \ref{total} one can see that, under the initial assumption $T_{\rm C} \ll T_{\rm M} \leq N$, 
the overall conductance is determined by the bare value $T_{\rm C}$ calculated above. On the other hand, 
in the absence of scattering
at both metal-ribbon interfaces and constriction, one recovers the perfect conductance of the ribbon. 
The function $T_{\rm M}$ is, unfortunately, hard to determine and is 
expected to be strongly dependent on the metal used. 
Experiments\cite{Heinze:prl:02,Javey:nature:03,Yang:apl:05,Chen:nl:05}  as well as first-principles calculations of this
quantity for carbon nanotubes\cite{Palacios:prl:03,Palacios:07}, have revealed
that $T_{\rm M}/N$ can range from 10$^{-3}$ to 1. For Pd electrodes, 
which have proved to make excellent contacts, $T_{\rm M}/N \approx 0.1-1$\cite{Javey:nature:03,Palacios:07}. For other
metals this factor can be reduced in orders of magnitude. To illustrate the effect of the metal on the performace of 
the GRFET we consider an electrode model in the form of a bidimensional 
square lattice with the same hopping parameter $t$ as the graphene lattice and varying on-site energy, $\epsilon$.  
The inset in Fig.  \ref{T300} shows the conductance of a single metal-graphene interface for various $\epsilon$.
As an illustration, the inset in Fig.  \ref{T300} shows $G_{\rm T}$ for the square constrictions using $\epsilon=5t$.
The performance of the GRFET's is noticeably
affected in all cases since $T_{\rm M} \approx T_{\rm C} $ close to $V_{\rm th}$.
Scattering at the metal is, however, less effective in reducing
$T_{\rm M}$ down to $\approx 1$ close to $V_{\rm th}$ as $W_{\rm C}/W_{\rm L}\rightarrow 0$. All this results 
in the following.  Scattering at the constriction increases as $W_{\rm C}/W_{\rm L}\rightarrow 0$.
On the other hand, the number of bulk ribbon channels at $V_{\rm th}$
also increases as $W_{\rm C}/W_{\rm L}\rightarrow 0$, reducing the pervasive effect of the contact resistance to
the metallic electrodes. One can thus conclude that an optimal ratio $W_{\rm C}/W_{\rm L}$ will always exist for GRFET's
and that this ratio will depend on both the metal contact and type of constriction.

{\it Final remarks.--}
When compared to bidimensional non-relativistic 
electrons\cite{Szafer:prl:89}, the scattering at the constriction is much stronger in the case of relativistic ones.
This prevents conductance quantization from appearing in these devices.
For tapered interfaces, deviations from conductance quantization can still be of the order of 2 or 3\%  while they
are of the order of 0.0001\% for non relativistic electrons\cite{Szafer:prl:89}. Our results support recent claims
by Katsnelson\cite{Katsnelson:07}, although more work is needed here to clarify these fundamental issues. 
When the constriction is placed symmetrically in the middle of the ribbon, the scattering at the contact turns out to be
larger, strongly worsening conductance quantization and the performace of the GRFET. 
A detailed account of this and other possible contact models that
can be considered will be published elsewhere\cite{Munoz-Rojas:07}.

Concerning disorder effects, a caveat should be issued. The use of 
of graphene ribbons for field effect transistors could be hampered by the fact that, in average, one out of three
constrictions are metallic, given the unlikely chance that lithography can define the constriction with atomic precision.
The same uncertainty in the lithographic definition of the constriction width, with the concommitant disorder at the edge,
could, however,  open up a gap, transforming otherwise metallic ribbons into a semiconducting ones\cite{Sols:07}.

In summary, our results  show, generically, that  the conductance of graphene ribbons with
constrictions presents a step-like behavior as a function of $E_{\rm F}$ or $V_{\rm G}$.  
However, one can hardly claim conductance quantization (in units of $2e^2/h$) in most cases.
At zero temperature, Fabry-Perot oscillations appear on top of the steps as a function of the gate voltage,
being more pronounced for the square constrictions. Such oscillations result from finite scattering at the interfaces,
which is reduced for the tapered ones, but not completely absent as naively expected.
At room temperature, the Fabry-Perot oscillations disappear
even for square interfaces and very short constrictions of $\approx 10$ nm.  Due
to the relativistic nature of electrons in graphene, $G_{\rm on}$
does not saturate to the quantum limit $2e^2/h$ close to $V_{\rm th}$ 
for neither type of constrictions considered. However,
from an operational point of view and compared to making direct contact to metal electrodes, one could say that 
tapered contacts between a graphene ribbon and a side constriction 
can perform nearly as ohmic contacts for intermediate constriction widths.

{\em Acknowledgements.--} F.M.R. is endebted to D. Jacob for his help with the transport code
and acknowledges the University of Alicante for financial support. 
We acknowledge F. Guinea and P. Jarillo-Herrero for discussions. 
This work has been funded by Spanish MEC under grants FIS2004-02356 and MAT2006-03741,
by Generalitat Valenciana under grant ACOMP07/054, and by FEDER Funds.
\\


\end{document}